\documentclass[twocolumn,prl,citeautoscript,showpacs]{revtex4-1}

\usepackage{hyperref}
\usepackage{natbib}

\usepackage{amssymb}
\usepackage{amsmath}
\usepackage{amsfonts}
\usepackage{array}

\usepackage{graphicx}

\newcommand{\ocite}[1]{[\onlinecite{#1}]}

\newcommand{\lbrs}{\left[}
\newcommand{\rbrs}{\right]}
\newcommand{\lbrc}{\left\{}
\newcommand{\rbrc}{\right\}}

\newcommand{\ten}[1]{\textsf{#1}}
\newcommand{\mat}[1]{\mathbb{#1}}
\renewcommand{\vec}[1]{\boldsymbol{#1}}

\newcommand{\beq}{\begin{eqnarray}}
\newcommand{\eeq}{\end{eqnarray}}
\renewcommand{\d}{{\textrm{d}}}
\newcommand{\Tr}{{\textrm{Tr}}}

\newcommand{\erf}{\textrm{erf}}

\newcommand{\half}{\frac{1}{2}}
\newcommand{\aeq}{\approx}

\newcommand{\Dp}[1]{\partial_{{#1}}}
\newcommand{\Da}{\Dp{\alpha}}

\newcommand{\eref}[1]{equation \eqref{#1}}

\newcommand{\erefs}[2]{equations \eqref{#1}-\eqref{#2}}

\newcommand{\Fref}[1]{Figure \ref{#1}}

\newcommand{\Tref}[1]{Table \ref{#1}}

\newcommand{\vdW}{\textrm{vdW}}

\newcommand{\chih}{{\hat{\chi}}}

\newcommand{\epsilonb}{{\bar{\epsilon}}}

\newcommand{\comment}[1]{}

\newcommand{\pr}{^{\prime}}

\renewcommand{\vr}{\vec{r}}
\newcommand{\vrp}{\vr\pr}

\newcommand{\vk}{\vec{k}}

\newcommand{\vq}{\vec{q}}

\newcommand{\vrpa}{\vec{r}_{\parallel}}

\newcommand{\qpa}{q_{\parallel}}

\newcommand{\vqpa}{\vec{q}_{\parallel}}

\newcommand{\vu}{{\vec{u}}}

\newcommand{\n}{n^0}
\newcommand{\N}{N^0}
\newcommand{\T}{\bar{T}^0}
\newcommand{\tT}{\bar{\ten{T}}^0}

\newcommand{\tKh}{\hat{\ten{K}}}
\newcommand{\Kh}{\hat{K}}

\newcommand{\bra}[1]{\left<#1\right|}
\newcommand{\ket}[1]{\left|#1\right>}

\newcommand{\rcite}[1]{ref.~\onlinecite{#1}}

\newcommand{\Rcite}[1]{Ref.~\onlinecite{#1}}
\newcommand{\Rcites}[1]{Refs.~\onlinecite{#1}}

\newcommand{\KS}{{\textrm{KS}}}
\newcommand{\dRPA}{{\textrm{dRPA}}}
\newcommand{\xc}{{\textrm{xc}}}
\newcommand{\xt}{{\textrm{x}}}
\newcommand{\ct}{{\textrm{c}}}
\newcommand{\lr}{{\textrm{lr}}}

\newcommand{\Ec}{E_{\ct}}
\newcommand{\Ex}{E_{\xt}}

\newcommand{\qRS}{q_{\textrm{RS}}}

\newcommand{\TV}{{\textrm{CM}}}

\newcommand{\TVF}[1]{{\TV_{(#1)}}}

\newcommand{\LDA}{{\textrm{LDA}}}

\newcommand{\LSR}{{\textrm{Lsr}}}
\newcommand{\EXX}{{\textrm{EXX}}}

\newcommand{\NEig}{N_{\textrm{Eig}}}
\newcommand{\NBas}{N_{\textrm{Bas}}}

\begin{document}
\title{Efficient, long-range correlation from occupied wavefunctions only}
\date{\today}
\author{Tim Gould}
\email[Email: ]{t.gould@griffith.edu.au}
\author{John F. Dobson}
\affiliation{QLD Micro- and Nanotechnology Centre, %
Griffith University, Nathan, QLD 4111, Australia}
\begin{abstract}
Via continuum mechanics [PRL 103,086401] with
Random Phase Approximation (dRPA) screening, 
we develop a numerically efficient general-geometry
electronic exchange-correlation 
energy functional. It gives correct asymptotic power laws
for dispersion interactions between
 insulators or metals. As a numerical example we obtain
the full binding energy curves 
$\epsilonb(D)$ for parallel metal slabs of small but
finite thickness: at all separations $D$ our 
$\epsilonb(D)$ agrees better with full dRPA correlation
calculations than does the Local Density 
Approximation, while being much more efficient than full dRPA correlation.
\end{abstract}

\maketitle

An increasing body of work\cite{*[{See page 1898 of }]vdWReview2010-Short,%
Harl2008,*Harl2009,Lebegue2010,Eshuis2010,*Eshuis2011}
has demonstrated that the correlation energy $\Ec^{\dRPA}$ in
the direct Random-Phase Approximation (dRPA) is highly accurate
for energy differences in many and varied electronic systems,
at least in cases where orbital self interaction is not an issue.
dRPA binding properties for a wide variety of bulk
materials\cite{Harl2008,*Harl2009} are typically more accurate
than those from the local density approximation (LDA), especially
for dispersion (van der Waals, vdW) bound systems\cite{Lebegue2010}.
For the vdW attractive potential, which is totally neglected in the
LDA, the dRPA proves to be versatile, predicting unusual
vdW coefficients\cite{Kim2006,*Cole2009}
and power laws\cite{Sernelius1998,*Dobson2006,*Gould2008,*Gould2009}
in agreement with quantum Monte Carlo results\cite{Drummond2007}.

$\Ec^{\dRPA}$ is typically obtained in three steps:
i) The bare response $\chih_0$ is obtained from occupied
and unoccupied groundstate wavefunctions.
This is typically the numerical bottleneck.
Recent developments\cite{Lu2009,*Nguyen2009} attempt to bypass
unoccupied states but can encounter problems for metallic systems.
ii) The interacting response is calculated through the dRPA as
$\chih_{\lambda}(\omega)=\chih_0(\omega) 
+ \lambda \chih_0(\omega) \hat{v}\chih_{\lambda}(\omega)$
where $\hat{v}$ is the Coulomb potential $|\vr-\vrp|^{-1}$.
iii) Finally the correlation energy is calculated via
integration on the imaginary frequency axis through
the Adiabatic Connection and Fluctuation Dissipation Theorem (ACFD)
approach
\begin{align}
\Ec^{\dRPA}=&-\int_0^{\infty}\frac{\d\sigma}{2\pi}
\Tr\lbrs \log\{\hat{1}+\hat{A}(i\sigma)\}-\hat{A}(i\sigma) \rbrs
\label{eqn:ACFDT}
\end{align}
where $\hat{A}(\omega)=-\hat{v}^{1/2}\chih_{0}(\omega)\hat{v}^{1/2}$
is an Hermitian operator\footnote{%
We adopt the following notations: $s$ is a scalar,
$\vec{v}$ is a 3D vector, $\ten{T}$ is a 3x3 tensor and
$\mat{M}$ is a general matrix.
Operators wear a hat $\hat{o}$.
Hartree atomic units with
$\hbar=e^2/(4\pi\epsilon_0)=m=1$ are used throughout. Greek subscripts
represent cartesian components and are summed over if repeated.
Derivatives $\Da$ and $\nabla$ act on \emph{everything}
to the right except in square brackets.}.

Other efficient van der Waals (vdW) functionals%
\cite{Grimme2006,*Tkatchenko2009,*Vydrov2009,%
Dion2004,*Rydberg2000,*Rydberg2003,*Langreth2005}
give good results for many systems. However they represent
$\Ec^{\vdW}$ in an additive two-point approximation that is either
obtained semi-empirically\cite{Grimme2006,*Tkatchenko2009,*Vydrov2009}
or derived\cite{Dion2004,*Rydberg2000,*Rydberg2003,*Langreth2005} by
solving the dynamical screening problem \eqref{eqn:ACFDT} perturbatively.
As a result, these functionals miss non-pairwise-additive
vdW energy contributions that can be substantial in
highly polarizable, highly anisotropic systems%
\cite*{Kim2006,*Cole2009,Sernelius1998,Dobson2006,*Gould2008,*Gould2009},
including low-dimensional metals.
Very large, anisotropic molecules
and metallic and graphitic surface physics (e.g. binding of graphite
on metal surfaces) are two classes
of systems where standard methods are inaccurate\cite{Vanin2010}
and dRPA is intractable.

Here we solve \eref{eqn:ACFDT} accurately thus avoiding the pairwise
additive approximation, but we use the continuum mechanics of
Tokatly, Tao, Gao and Vignale%
\cite*{Tokatly2007,Tao2009,*Gao2010}
to approximate $\chih_0$ in a numerically efficient manner.
Their linearized continuum mechanics (CM) scheme \cite{Tao2009}
uses the continuum fluid displacement $\vu$,
which is related to the density perturbation $n^1$
by\cite{Dobson1994,Tao2009,*Gao2010}
\begin{align}
n^{1}(\vr,t)=-\Dp{\mu} [\n(\vr)u_{\mu}(\vr,t)].
\label{eqn:deln}
\end{align}
For a small change to the Kohn-Sham (KS) potential
$V^{1}(\vr,t)$ CM theory approximates 
$\vu$ through the following hydrodynamic-like equation
(from equations 3, 4 and 14-16 of \ocite{Tao2009})
\begin{align}
\Dp{tt} u_{\mu}(\vr,t) =& 
\frac{-\Phi^0_{\mu\nu} u_{\nu}(\vr,t) + F^0_{\mu}(\vr,t)}{\n(\vr)}
-\Dp{\mu} V^{1}(\vr,t)
\label{eqn:TVMain}
\end{align}
where $\n(\vr)$, $\Phi^0_{\mu\nu}=-\n(\vr)[\Dp{\mu\nu} V^{\KS}(\vr)]$
and $F^0_{\mu}(\vr,t)$ depend on groundstate properties of the system.

The force $F^0_{\mu}$ is defined in equation 14 of \ocite{Tao2009}.
Careful manipulation of equation 14 allows us to write
it as $F^0_{\mu}=-\Kh_{\mu\nu}u_{\nu}(\vr,t)$. Here $\tKh$
is a tensor, Hermitian ($\Kh_{\mu\nu}=\Kh^{\dag}_{\nu\mu}$) operator
defined by
\begin{align}
\Kh_{\mu\nu}=&\Kh_{\mu\nu}^{(T)} - \frac{1}{4}\Kh_{\mu\nu}^{(n)}
\label{eqn:KDef1}
\allowdisplaybreaks\\
\Kh_{\mu\nu}^{(T)}=& \Dp{\alpha}\T_{\mu\nu}\Dp{\alpha}
+ \Dp{\nu}\T_{\mu\alpha}\Dp{\alpha} + \Dp{\alpha}\T_{\alpha\nu}\Dp{\mu}
\label{eqn:KDef2}
\allowdisplaybreaks\\
\Kh_{\mu\nu}^{(n)}=&
\Dp{\nu\alpha} \n(\vr) \Dp{\alpha\mu}.
\label{eqn:KDef3}
\end{align}
It involves the electron density and
groundstate kinetic stress tensor $\T_{\mu\nu}=
\Re \sum_i f_i [\Dp{\mu}\psi_i(\vr)]^*[\Dp{\nu}\psi_i(\vr)] -
\frac{[\Dp{\mu\nu} \n(\vr)]}{4}$ %
\footnote{We define the kinetic stress tensor $\tT$ slightly
differently from Tao \emph{et al} but the force is identical}
where the sum is over occupied orbitals.

In the absence of an external potential, \eref{eqn:TVMain}
has time-periodic eigen-solutions defined by the hydrodynamic
eigen-equation
\begin{align}
-\Omega_N^2 \n u_{N\mu}(\vr) =&
[\Phi^0_{\mu\nu} + \Kh_{\mu\nu}] u_{N\nu}(\vr)
\label{eqn:TVEig}
\end{align}
where $N$ labels the sorted eigen-modes,
$\vu_N(\vr)$ is related to an
eigen-function of $\chih_0$, $\Omega_N>\Omega_{N-1}$ is related
to the KS excitation energies (exactly in one-electron systems) and
$\int\d\vr \n(\vr) \vu_N^*(\vr)\cdot\vu_M(\vr)=\delta_{NM}$.

By definition the tensor polarizability 
$X_{0\mu\nu}(\vr,\vrp;\omega)$ is the time-periodic response
of the $\mu$ cartesian component of the 
polarization $-\n(\vr) \vu_{N} (\vr)$ to an
external electric field in the $\nu$
direction, while 
$\chi_0$ is the change in density $n^1(\vr)$ in response
to a small change in the
KS potential of form $V^1(\vr;\vrp)=\delta(\vr-\vrp)$.
They can be obtained through \erefs{eqn:deln}{eqn:TVMain}
and expansion in the eigen-solutions of \eqref{eqn:TVEig}
provides the convenient forms (where in practise $N$ is summed over
the lowest $\NEig$ eigen-pairs)
\begin{align}
X_{0\mu \nu }(\vr,\vrp,i\sigma)=
\sum_{N}F_{N}(i\sigma)p^*_{N\mu }(\vr)p_{N \nu }(\vrp)
\label{eqn:AlphaDef}
\allowdisplaybreaks \\
\chi_{0}(\vr,\vrp,i\sigma)=-\sum_{N}F_{N}(i\sigma)d^*_{N}(\vr)d_{N}(\vrp)\;\;
\label{eqn:ChiDef}
\end{align}
where $F_N(i\sigma)=(\Omega_N^2+\sigma^2)^{-1}$,
$\vec{p}_N = \n(\vr)\vu_N(\vr)$ and
$d_N(\vr)=[-\nabla \cdot \{\n(\vr) \vu_N(\vr)\}]=-[\nabla\cdot\vec{p}_N]$.

For efficient evaluation of the correlation energy,
the following important relationships
are derived which allow us to evaluate $\Ec$ using integrals
over \emph{one space variable} only, reducing calculation time
and storage requirements%
\footnote{A similar approach is often employed
for exact dRPA of molecules.
It is usually impractical in periodic bulk systems due to
the need for $O(N_{\vk})$ Brillioun zone transitions
(with $N_{\vk}$ points sampled). The CM avoids this issue.}.
From \eqref{eqn:ChiDef} the projection of $\hat{A}$
[see \eqref{eqn:ACFDT}] in reciprocal space can be written in
the separable form
$A(\vq,\vq')=\sum_N F_N(i\sigma) w_N^*(\vq)w_N(\vq')$
where
\begin{align}
w_N(\vq)=&
-i\vq\sqrt{v(q)} \cdot \int \d\vr e^{i\vq\cdot\vr} \n(\vr) \vu_N(\vr)
\label{eqn:wN}
\end{align}
or $w_N=\sqrt{v(q)}d_N(\vq)$ (here $v(q)=4\pi q^{-2}$).
Setting $W_{NM}=\int \frac{\d\vq}{(2\pi)^3} w_N(\vq) w_M^*(\vq)$
allows us to define an $\NEig\times\NEig$ matrix
$\mat{B}(i\sigma)$ with elements
\begin{align}
B_{NM}(i\sigma)=&\sqrt{F_N(i\sigma)F_M(i\sigma)} W_{NM}
\label{eqn:BNM}
\end{align}
with $\lim_{\NEig\to\infty}\Tr_N[G(\mat{B}(i\sigma))]
=\Tr_{\vq}[G(\hat{A}(i\sigma))]$\cite{Supp}
for any analytic function $G$.

Finally, defining the eigen-values of $\mat{B}(i\sigma)$ to be
$\beta_{\kappa}(i\sigma)$ we reduce the correlation energy \eqref{eqn:ACFDT}
to the form
\begin{align}
\Ec^{\TV}
=&-\int_0^{\infty} \frac{\d\sigma}{2\pi} \sum_{\kappa}
\lbrc \log[1+\beta_{\kappa}(i\sigma)] - \beta_{\kappa}(i\sigma) \rbrc.
\end{align}
In practice we seem only to need a small number
$\NEig$ of eigen-solutions to converge correlation energies to a
sufficiently small error ($\propto 1/\Omega_{\NEig}^3$)
within CM theory\cite{Supp}.
This agrees with other observations (e.g. \rcite{Nguyen2009})
that calculating $\Ec$ through a
diagonalisation of $\chih_0\hat{v}$ requires few eigenvalues
for convergence.

The most trying calculation in this functional method is evaluation of
\eref{eqn:TVEig}, as $\tKh$ is a spatially-dependent,
differential operator. To overcome this problem
we use an auxiliary basis set $\mathcal{B}\equiv \lbrc\phi_j(\vr)\rbrc$,
of size $\NBas$,
which need not be mutually orthogonal but must be complete in the
limit $\NBas\to\infty$.  
Choice of this basis is the only part of his scheme that differs for different 
geometries or systems: for example, plane waves for periodic systems, %
gaussians for atoms and molecules.
With a given basis set we expand our CM eigen-function \eqref{eqn:TVEig} as
$u_{N\mu}(\vr)=\sum_j a_{N\mu}^j \phi_j(\vr)$
which we substitute into \eref{eqn:TVEig}. This provides
a set of $3\NBas\times 3\NBas$ coupled equations
\begin{align}
-\Omega_N^2 \N_{jk} a_{N\mu}^k =& \lbrc \Phi^0_{jk\mu\nu} + K_{jk\mu\nu} \rbrc
a_{N\nu}^k
\label{eqn:TVBasis}
\end{align}
while $\N_{jk} (a_{N\mu}^{j*}a_{M\mu}^k)=\delta_{NM}$ sets the orthogonality.

The non-operator terms in these equations are
$\N_{jk}=\int \d\vr  \n(\vr)\phi_j^*(\vr)\phi_k(\vr)$
and $\Phi^0_{jk\mu\nu}=-\int \d\vr [\n(\vr)\Dp{\mu\nu}V^{\KS}(\vr)]
\phi_j^*(\vr)\phi_k(\vr)$. Separating the final term into
$K_{jk\mu\nu}=\int \d\vr \phi_j^*(\vr) \Kh_{\mu\nu}(\vr) \phi_k(\vr)
=K^{(T)}_{jk\mu\nu} - \frac{1}{4}K^{(n)}_{jk\mu\nu}$
and using integration by parts 
gives
\begin{align}
K^{(T)}_{jk\mu\nu} =& -\int \{
\T_{\mu\alpha}[\Dp{\nu}\phi_j^*][\Dp{\alpha}\phi_k]
+ \T_{\alpha\nu}[\Dp{\alpha}\phi_j^*][\Dp{\mu}\phi_k]
\nonumber\\&\hspace{10mm}
+ \T_{\mu\nu} [\nabla\phi_j^*]\cdot[\nabla\phi_k]\}d\vr
\label{eqn:BasisTerms1}
\\
K^{(n)}_{jk\mu\nu} =& \int
\n [\Dp{\nu}\nabla\phi_j^*]\cdot[\Dp{\mu}\nabla\phi_k]
\d\vr
\label{eqn:BasisTerms3}
\end{align}
where all terms are functions of $\vr$ and all
derivatives can, ideally, be performed \emph{analytically}
on the basis functions.

Surprisingly for a hydrodynamic-style approach, CM theory gives
the exact bare responses  $\ten{X}_0$, $\chi_{0}$ to
irrotational fields of one- and two($\uparrow\downarrow$)-electron
systems around their groundstate\cite{Tokatly2007,Supp}.
This means that our correlation scheme will give the same results 
as dRPA for the asymptotic vdW interaction
between two hydrogen or two helium atoms.

To explore this further we follow \cite{Gao2010}
in expanding both the KS and CM response to $O(\omega^{-4})$
leading to the following identities\cite{Supp}
\begin{align}
1=&\sum_{ja} h_{jaN},&
\Omega_N^2=&\sum_{ja} h_{jaN}\omega_{ja}^2.
\label{eqn:SumRule}
\end{align}
Here $h_{jaN}=\frac{2|f_j-f_a||K_{jaN}|^2}{|\omega_{ja}|}$
where $\omega_{ja}$ is a KS eigen-energy difference of an
occupied orbital $\ket{j}$
and unoccupied orbital $\ket{a}$ with occupations $f_j$ and $f_a$, and
$K_{jaN}=\int\bra{j}\hat{\vec{J}}\ket{a}\cdot\vu_N\d\vr $ is a mode-overlap
matrix element of the current (obtained via the current operator
$\hat{\vec{J}}$).
Thus $\Omega_N\geq\epsilon_{\rm{L}}-\epsilon_{\rm{H}}$
for isolated systems and
$\Omega_{N\vq}\geq\min_{\vk}(\epsilon_{\rm{L}\vk+\vq}-\epsilon_{\rm{H}\vk})$
for periodic systems where L labels the lowest unoccupied orbital or band
and H labels the highest occupied.

One implication of this is that a Kohn-Sham insulator
will remain an insulator under CM, in the sense of finite responses
\eqref{eqn:AlphaDef}-\eqref{eqn:ChiDef} as $\sigma\to 0$.
Thus\cite{Supp} CM theory obeys the well-known vdW laws for insulators
with (e.g.) a $-C_4D^{-4}$ asymptotic binding for two thin layers.
This is a very strong feature of the CM theory, not shared by common
approximated ACFD theories
\cite{Dion2004,*Rydberg2000,*Rydberg2003,*Langreth2005,Dobson1999}
where explicit cutoffs have to be imposed in
the tails in order to suppress metallic-like response.

In the opposite limit of a homogeneous electron gas (HEG),
CM is analytically soluble,
agrees with the true $\chi_{0}$ for $q\ll k_F$, $\omega \gg v_{F}q$,
and in particular has a ``metallic'' infinite polarizability,
$\ten{X}_0 \rightarrow \infty $ as $q$ and $\omega\rightarrow 0$.
Electron-gas-like (metallic) systems nevertheless pose
a difficult test for CM theory because the single-particle-like
excitations occurring for $\omega <v_{F}q,$ (and thus not
accurately desribed by CM), can make significant contributions to
the RPA correlations, mainly at short spatial range
(large wavelength).

This inaccuracy can be improved in metallic systems
by employing range-separation (RS)
such that the short-range physics is treated by a local scheme.
This makes \emph{no contribution} to vdW asymptotic physics.
A well-studied RS scheme is described in
\cite{Savin1995,*Leininger1997,*Gerber2005}.
It involves choosing a $\qRS$ and splitting up the Coulomb
potential, with a long-range component $v^{(\qRS)}(r)=\erf(\qRS r)r^{-1}$,
equivalent to replacing \eqref{eqn:wN} by
$w_N^{(\qRS)}(\vq)=w_N(\vq)e^{-q^2/(8\qRS^2)}$. We label the corresponding
correlation energy $\Ec^{\lr\TVF{\qRS}}$. This has the additional
benefit of accelerating convergence.

For $\chih_0$ to be reliably approximated
by continuum mechanics without a separate treatment of the low
frequencies we must choose $\qRS$ to
be \emph{substantially} less than $k_F$. Here we use
$\qRS=0.25 r_s^{-1}=0.13 k_F$
where $r_s$ is a global measure of the inter-electron distance.
For the jellium slab problems studied below we simply choose $r_s$
corresponding to the background charge density of each slab,
though more general prescriptions exist.
The remaining correlation must be included from local approximations
so that
\begin{align}
  \Ec^{\TVF{\qRS}}[n]=& \Ec^{\lr\TVF{\qRS}}
  + \int\d\vr n(\vr)\epsilon_{\ct}^{\LSR{(\qRS)}}
\label{eqn:EcTV}
\end{align}
where $\epsilon_{\ct}^{\LSR{(\qRS)}}$ is the correlation energy per electron
of the HEG with a short-ranged interaction, taken from\cite{Toulouse2004}.

Ideally we must also implement a range-separation for exchange,
but this proves numerically difficult for the slab geometries
we investigate. We instead use the ratio of the long-range exchange
to total exchange of an HEG
$A_{\xt}\aeq 1.1\qRS r_s/\sqrt{1+(1.1\qRS r_s)^2}$
as a prefactor for the exact exchange (EXX)
$\Ex^{\EXX}=-\half\int\d\vr\d\vrp|\vr-\vrp|^{-1}\allowbreak
|\sum_n f_n\psi_n^*(\vr)\psi_n(\vrp)|^2$
and make up the remainder with the LDA.
Combining this with \eqref{eqn:EcTV} gives
$E_{\xc}=A_{\xt} \Ex^{\EXX} + (1-A_{\xt}) \Ex^{\LDA} + \Ec^{\TVF{\qRS}}$.

As a numerical test of our proposed functional we choose the difficult
case of two thin metal slabs
described in \Rcites{Dobson1999,Dobson2004,Jung2004}.
This system is defined by three parameters only:
the width of the slabs $s$, the inner surface-surface distance $D$
and the positive background charge electron density
$\rho=3/(4\pi r_s^3)$.
The total number of electrons per unit area
is $N_s=2s\rho=\int_{-\infty}^{\infty}n^+(z)\d z$.

\begin{table}[b]
\begin{tabular*}{0.95\columnwidth}{@{\extracolsep{\fill}}|l|rrr|rrr|}
\hline\hline
& CM & LDA & dRPA
& CM & LDA & dRPA
\\ \hline
& \multicolumn{3}{c|}{$r_s=1.25$, $s=3$}
& \multicolumn{3}{c|}{$r_s=2.07$, $s=5$}
\\ \hline
$D_0$ & 3.33 & 3.38 & 3.32\ddag & 1.57 & 1.56 & 1.62$\pm$0.1\S
\\
$\epsilon_b$ & 0.74 & 0.53 & 0.79\ddag & 1.78 & 1.72 & 1.85$\pm$0.1\S
\\
$C_{zz}$ & 0.51 & 0.45 & 0.55\ddag & 1.31 & 1.38 & 1.32$\pm$0.1\S
\\\hline\hline
\end{tabular*}
\caption{Groundstate properties of two slab systems
under different approximations.
Energies are in mHa/$e^-$ and distance are
in Bohr radii.
\ddag\ from \Rcite{Jung2004}, \S\ is guessed from \Rcites{Dobson1999,Dobson2004}
taking into account estimated error bars. %
\label{tab:Binding125}}
\end{table}
\begin{figure}[t]
\includegraphics[width=0.95\linewidth]{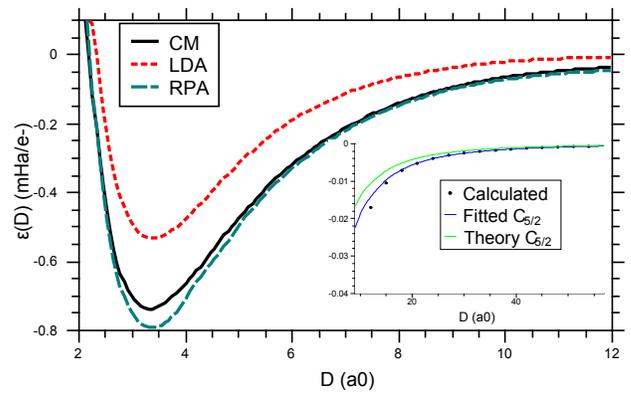}
\caption{$\epsilonb(D)$ graph for $r_s=1.25$, $s=3$.
RPA data from \cite{Jung2004}. Inset data shows the vdW dominated region.
\label{fig:Plot1}}
\end{figure}
We test our method on slab pairs with
$s=3a_0$, $r_s=1.25a_0$ and $s=5a_0$, $r_s=2.07a_0$ which have been
studied in \Rcite{Jung2004} and \Rcites{Dobson1999,Dobson2004} respectively.
Especially in the first case the LDA and dRPA give
significantly different energy curves.
We consider the cleavage energy per electron
$\epsilonb(D)=\epsilon(D)-\epsilon(\infty)
=[E_0(D)-E_0(\infty)]/N_s$ as a function of $D$.
Slabs with $r_s<4$ have a defined binding length $D_0$
where the force is zero. Thus a binding energy
$\epsilon_b=|\epsilonb(D_0)|$ and an elastic modulus
$C_{zz}=\Dp{DD}\epsilonb(D_0)$ can also be defined.

In \Fref{fig:Plot1} we plot $\epsilonb(D)$ versus $D$
for $r_s=1.25$, $s=3$. Our method matches the RPA closely
for this system. Binding properties for both studied systems
are tabulated in \Tref{tab:Binding125}
and show that the $r_s=2.07$, $s=5$ system is less well-predicted
but still much better than the LDA. If instead we set $\qRS=\infty$ the
results become much worse for both cases.
For widely separated slabs ($D\gg s$) the CM theory 
correctly and analytically describes
coupled two-dimensional plasmons
and hence correctly predicts the known asymptotic dRPA form\cite{Sernelius1998}
$\epsilonb(D\gg s)\aeq -0.012562\sqrt{N_s}(D+s)^{-5/2}$.
With $s=12.8a_0$ and $r_s=2a_0\ldots 6a_0$ we calculate
$C_{5/2}$ numerically within 8\% of the theory.
By contrast most other efficient vdW functionals would
predict an incorrect power law exponent in this limit with
$\epsilonb(D)\aeq -C_4D^{-4}$.

%
In our CM calculations we use auxiliary
basis functions $\phi_{k\vqpa}(\vr)=b_k(z)e^{-i\vqpa\cdot\vrpa}$\cite{Supp}
at $N_{\qpa}$ $\vqpa$ points.
All calculations are quite efficient with the slowest step being
evaluation of $W_{NM}$ at $O(N_{\qpa}N_{q_z}\NBas^2)$.
Convergence is reached with $\NBas=42$, $N_{\qpa}=55$,
$N_{\sigma}=250$ and $\NEig\leq 60$. Our dRPA calculation
takes approximately eight times longer than the groundstate
LDA calculation.
Test runs of full RPA calculations for these systems took hours,
compared to minutes for our functional, consistent with $N_{\qpa}=55$.
We also note that a 10\% variation in $\qRS$
made only a 1\% change to $\epsilon_b$.

While results for our test systems are not perfect, they show
closer agreement with the dRPA than the LDA both in the binding
region and for larger $D$, with a marked improvement
in speed over full dRPA. The vdW dispersive
physics is treated accurately and shows excellent agreement
with the dRPA in contrast to other methods.
The current prescription has a wide scope for refinement
both empirically through adjustment of $\qRS$ and the exchange functional
and by introducing better physics, most obviously
through improved (semi-local) treatment of low-frequency
behaviour which will reduce dependence on the range separation.

Furthermore preliminary tests suggest 
that metals are a worst-case for CM theory - i.e. that range separation 
will be much less needed for bound and insulating systems.

In summary, we have derived and developed an
\emph{efficient} general-geometry functional with correct
long-range correlation energy. Its ability to predict
correctly the vdW physics of metallic and insulating systems
is a distinct advantage over other efficient vdW functionals.
It is currently being implemented for periodic systems, which should 
enable meaningful energy calculations for (e.g.)
vdW bonded nanosystems such as metallic nanotube arrays
or graphene on metals. These systems require non-pair-additive 
high-level computations (e.g. dRPA) with a large unit cell, which
is beyond present computational power.  

\begin{acknowledgments}
The authors would like to thank I. Tokatly, J. Jung, A. Savin, J. \`Angy\`an, ,
and G. Vignale for fruitful discussions.
\end{acknowledgments}

\bibliography{vanDerWaals,ACFDT,DFT,Wannier,Misc,Experiment,Hybrid,CM-RPA}

\end{document}


\title{Supplementary material for %
``Efficient, long-range correlation from occupied wavefunctions only''}
\date{\today}
\author{Tim Gould}
\email[Email: ]{t.gould@griffith.edu.au}
\author{John F. Dobson}
\affiliation{QLD Micro- and Nanotechnology Centre, %
Griffith University, Nathan, QLD 4111, Australia}

\maketitle

\section{Equivalence of traces}

The calculation of $E_c$ via $\ten{W}$ relies on an equivalence
of certain traces. Here we demonstrate that
\begin{align}
\Tr_{\vq}[\log(1+\hat{A})-\hat{A}]=&\Tr_{N}[\log(\mat{I}+\mat{B})-\mat{B}]
\end{align}
where $\hat{A}=-\hat{v}^{\half}\chih_0\hat{v}^{\half}$,
$B_{NM}=\sqrt{F_N(i\sigma)F_M(i\sigma)}W_{NM}$
and $W_{NM}=\int\dq{} w_N(\vq)w_M^*(\vq)$.

Let us project $\hat{A}$ onto Fourier space giving
\begin{align}
\hat{A}(\vq,\vqp)=&\sum_{NM} [\mat{F}]_{NM}w_N^*(\vq)w_M(\vqp)
\end{align}
where $F_{NM}=F_N(i\sigma)\delta_{NM}$ and we subsequently omit $\sigma$
for notational brevity.
We can then evaluate
\begin{align}
\hat{A}^2(\vq,\vqp)=&\int\dq{2}\hat{A}(\vq,\vq_2)\hat{A}(\vq_2,\vqp)
\nonumber\\
=&\int\dq{2}
\sum_{N_1M_1} [\mat{F}]_{N_1M_1}w_{N_1}^*(\vq)w_{M_1}(\vq_2)
\nonumber\\&\times
\sum_{N_2M_2} [\mat{F}]_{N_2M_2}w_{N_2}^*(\vq_2)w_{M_2}(\vqp)
\nonumber\\
=&\sum_{NM} w_N^*(\vq)w_M(\vqp)
[\mat{F}\mat{W}\mat{F}]_{NM}
\end{align}
or more generally
\begin{align}
\hat{A}\hat{M}(\vq,\vqp)=&-\sum_{NM} w_N^*(\vq)w_M(\vqp)
[\mat{F}\mat{W}\mat{M}]_{NM}
\end{align}
when $\hat{M}(\vq,\vqp)=\sum_{NM} w_N^*(\vq)w_M(\vqp) [\mat{M}]_{NM}$.

Observation of $\hat{A}^2$ suggests that
\begin{align}
\hat{A}^p(\vq,\vqp)=&\sum_{NM} w_N^*(\vq)w_M(\vqp)
[(\mat{F}\mat{W})^{p-1}\mat{F}]_{NM}.
\label{eqn:Ap}
\end{align}
Premultiplying by $\hat{A}$ validates this assumption since
\begin{align}
\hat{A}\hat{A}^p(\vq,\vqp)=&\sum_{NM} w_N^*(\vq)w_M(\vqp)
[\mat{F}\mat{W}(\mat{F}\mat{W})^{p-1}\mat{F}]_{NM}
\nonumber\\
=&\hat{A}^{p+1}(\vq,\vqp)
\end{align}
and thus \eqref{eqn:Ap} is the correct form by induction
on $p$.

Finally we can take traces (and use permutations under a trace)
to show
\begin{align}
\Tr_{\vq}[\hat{A}]=&\sum_N \int\dq{} w_N^*(\vq)w_N(\vq)[\mat{F}]_{NN}
\nonumber\\
=&\sum_N W_{NN}[\mat{F}]_{NN}
\nonumber\\
=&\Tr_{N}[\mat{F}\mat{W}]=\Tr_{N}[\mat{B}]
\\
\Tr_{\vq}[\hat{A}^p]=&\sum_{NM} \int\dq{} w_N^*(\vq)w_M(\vq)
[(\mat{F}\mat{W})^{p-1}\mat{F}]_{NM}
\nonumber\\
=&\sum_{NM} W_{MN}[(\mat{F}\mat{W})^{p-1}\mat{F}]_{NM}
\nonumber\\
=&\Tr_{N}[(\mat{F}\mat{W})^p]
=\Tr_{N}[\mat{B}^p]
\end{align}
where $\mat{B}=\sqrt{\mat{F}}\mat{W}\sqrt{\mat{F}}$ and
$B_{NM}=\sqrt{F_N(i\sigma)F_M(i\sigma)}W_{NM}$.
Finally it follows from Taylor expansion of an analytic function
$G(\hat{A})=\sum_p g_p\hat{A}^p$
that
\begin{align}
\Tr_{\vq}[G(\hat{A})]=&\sum_p g_p\Tr_{\vq}[\hat{A}^p]
=\Tr_{N}[G(\mat{B})].
\end{align}
Setting $G(x)=\log(1+x)-x$ gives the equivalence.

\section{Eigenvalue convergence}

In the main work we mention the rapid convergence with respect
to the CM eigen-pairs. While this is somewhat system specific
some general considerations make it likely that
fewer eigen-pairs than Kohn-Sham (KS)
transition modes will be required to
obtain the same convergence in most systems.

For any given KS problem we can expand solutions in a finite basis set
(eg. a real space grid, planewaves, Gaussians etc.)
of size $\NBas$ which can be made as large as we like.
The KS equations thus have $\NBas$ solutions of which $N_{\rm{Occ}}$
may be considered occupied and $\NBas-N_{\rm{Occ}}$ are unoccupied.
Since $\chih_0$ involves a sum over occupied and unoccupied states,
calculations involving $\chih$ have a leading
$O(N_{\rm{Occ}}(\NBas-N_{\rm{Occ}}))\aeq O(N_{\rm{Occ}}\NBas)$
if we assume $N_{\rm{Occ}}\ll \NBas$.

By contrast the tensor CM equations have $\NEig=3\NBas$ solutions.
Projection of $\chih_0$ involves a sum over all solutions and
takes $O(3\NBas)$. Combining this with the sum
rules [see (16) of main text] means that the CM eigen-frequencies
will be distributed more sparsely than the KS-transitions (except
in one and $\uparrow\downarrow$ two electron systems).

In the specific case of a periodic system this has a very
notable effect on integration over the Brillioun zone.
Here a natural basis set for both the LDA and CM is
$e^{i(\vq+\vG)\cdot\vr}$ where $\vq$ lies within the Brillioun zone
and $\vG$ is a reciprocal lattice vector.
It is thus sufficient to project $\chih_0$ onto
$\vq$, $\vG$ and $\vGp$. Following the standard notation
we use $\vk$ rather than $\vq$ for the groundstate sampling.

In a full KS-dRPA calculation, projection of $\chih_0$
is $O(N_{\rm{Occ\ band}}N_{\rm{Band}}N_{\vk}^2N_{\vG}^2)$ where
$N_{\rm{Occ\ band}}$ is the number of occupied bands and
$N_{\rm{Band}}$ is the total number of unoccupied bands.
$N_{\vk}$ is the number of points sampled in the Brillioun zone
and must be counted twice: once for each projection onto $\vq$
and once to cover all transitions from $\vk$ to $\vk+\vq$
(see e.g. section III of \rcite{Harl2008} for a more
comprehensive discussion).
By contrast the same projection in CM is
$O(\NEig N_{\vk}N_{\vG}^2)$ as all $\vk$ to $\vk+\vq$
transitions are treated collectively in a single $\vu_{N\vq}$ eigenmode.

This relationship can also be observed if we solve for $\chih$ by
direct perturbation\cite{Nguyen2009} instead of the diagonalization method
used here. In a full KS calculation we need to
perturb each of the $N_{\rm{Occ}}$ orbitals to obtain
$\chih$. In the CM we need only to solve for the
three-dimensional quantity $\vu$.
In the periodic bulk case $N_{\rm{Occ}}=N_{\vk}N_{\rm{Occ\ band}}$.

With a similar energy cutoff for both the KS and CM solutions
so that $\NEig\aeq N_{\rm{Band}}$, a planewave-based evaluation
of the CM-dRPA will run at least $O(N_{\vk})$ times quicker
than a full dRPA calculation.
In an efficient dRPA code in an insulating system, $N_{\vk}$
is typically found to be converged with
between 200 and 2000 points\cite{Harl2008} divided by
the number of symmetries.

Under the dRPA, the truncation error caused by using only a finite
number of transitions, is $O(1/\omega_{a_mj_m}^3)$ where
$\omega_{a_mj_m}=\epsilon_{a_m}-\epsilon_{j_m}$
(where $a_m$ is an occupied KS state and $j_m$ is unoccupied)
is the largest transition frequency
included in the sum. The equivalent error for the
CM is $O(1/\Omega_{N_m})$ where $\Omega_{N_m}$ is the transition
frequency of the highest included mode.

We must note, however, that there are a variety of ways to calculate
$E_c$. For example we can bypass the projection of $\chih_0$,
as we do in the method presented in the manuscript, by using
$\mat{W}$. For periodic bulks 
this is an $O(N_{\vk}\NBas^2N_{\vG})$ calculation under
the CM with $O(\NBas^2)$ storage and $O(\NBas^3)$ diagonalisation.
In the full KS it would be $O(N_{\vk}^3N_{\rm{Occ}}^2\NBas^2N_{\vG})$
with $O(N_{\vk}^2N_{\rm{Occ}}^2\NBas^2)$ storage 
and $O(N_{\vk}^3N_{\rm{Occ}}^3\NBas^3)$ diagonalisation
which is infeasible for most bulk systems.

As a general rule, projection of $\chih_0$ onto space will be
more efficient than using $\mat{W}$ 
for a true KS bulk or periodic system.
Calculation of $\mat{W}$ will be quicker for the CM in almost
all geometries and will be quicker than evaluation of $\mat{W}$
for the full KS system in all but the smallest systems.

In summary, for systems with more than a handful of occupied orbitals,
the CM is all but guaranteed to converge faster than the full KS
under the same method for calculating $E_c$.
Furthermore it may offer the potential to use alternate methods that are
faster still (such as using $\mat{W}$ for correlation energies) which
may be infeasible or much slower in the full KS system.

\section{Exactness in one-electron systems}

Equations (41), (45-46) of \Rcite{Tokatly2007} provide a proof that
the continuum mechanics approach is exact for one-electron
(or two-electron with equal spin up and down) systems. This follows
from the equivalence of CM to
Madelung dynamics which are equivalent to the Schr\"odinger
equation for one-orbital systems in irrotational fields\cite{Wallstrom1994}.
We will provide a direct proof from
the Schr\"odinger equation elsewhere.

\section{Non-contributing gauge modes}

It is possible to find solutions $\vu_N$ of the CM equations
where $\nabla\cdot\n\vu_N=0$. For example,
in the 2DEG systems examined modes of form
$\vu_N=\frac{c}{\n(z)}\vq_{\perp}e^{i\vqpa\cdot\vr}$ where
$\vq_{\perp}=\vqpa\times\hat{\vec{z}}$ have this property.

For such modes (labelled $N^*$) the weights of
the sum rules [equation (16) of main text]
become $h_{jaN^*}=0$. While a full proof is beyond
the scope of this supplement it follows from the exactness of the
CM response to $O(\omega^{-4})$
and the near-completeness over all occupied/unoccupied pairs $ja$
of $\bra{j}\hat{\vec{J}}\ket{a}$ for
vectors with non-zero gradients.
Thus $\sum_{ja} h_{jaN^*}=0$ and
$\Omega_N^2=\sum_{ja} h_{jaN^*}\omega_{ja}^2=0$.
One consequence of this is that any CM eigen-mode of
a KS-insulating system which has $\Omega_{N^*}=0$
must also have $\nabla\cdot\n\vu_{N^*}=0$.

Such modes will either be supressed by the boundary conditions
of the problem, or not contribute to the correlation energy
of the system and may thus be discarded. More precisely they do not
contribute to the linear response $\chih_0$ at all and contribute
only a shift of gauge to the tensor response $\ten{X}_0$.
The former identity can be seen directly as only $\nabla\cdot\n\vu_N$
appears in the expression [equation (9) of main text] for $\chih_0$
while the latter comes indirectly from
integration by parts over the Coulomb tensor.

\section{Interacting insulator}

We show in the main work that a KS insulating system (at a bare response
level) will remain insulating under the CM. More formally
this means that both the KS system and the CM approximation
thereto, obey
\begin{align}
|\tenh{X}_{0\vq}(\vG,\vGp;i\sigma)|\leq& \frac{Y}{W^2+\sigma^2}
\end{align}
where $Y$ and $W$ are undetermined but non-zero constants. In the CM
case this follows from $\vec{p}_N(\vr)=\n(\vr)\vu_N(\vr)<\infty$,
$\vec{p}_N(\vr\to\infty)=0$ and $\Omega_{N\vq}>0$ and the asymptote
$\lim_{\sigma\to\infty}|\tenh{X}_{0\vq}|=C/\sigma^2$.

Using the $\vec{p}_N$ and $\vec{p}_M$
matrix representation of the RPA-interacting response
$\tenh{X}_{\lambda}$ it is possible to show
\begin{align}
\tenh{X}_{\lambda\vq}(\vG,\vGp;i\sigma)=&
\sum_{H} \sum_{NM} \frac{X_{HNM}\vec{p}_{N\vq}^*(\vG)\vec{p}_{M\vq}(\vGp)}%
{\Omega_{\lambda H\vq}^2+\sigma^2}
\end{align}
where $\Omega_{\lambda H\vq}^2$ is an eigenvalue of
the matrix $\mat{D}_{\lambda \vq}=\rm{diag}[\Omega_{N\vq}^2] + \lambda \mat{W}_{\vq}$.
Provided $\mat{D}_{\lambda \vq}$ does not have any zero eigenvalues
(this would be a strongly correlated metallisation and is extremely rare)
it is obvious that
\begin{align}
|\tenh{X}_{\lambda\vq}(\vG,\vGp;i\sigma)|\leq \frac{Y'}{W'^2+\sigma^2}
\end{align}
for some finite $Y'$ and $W'$.

\section{Thin insulators have {$D^{-4}$} power laws}

If we have two well-separated electronic systems at
a distance $D$ with no electronic overlap between them, we can write the
dRPA dispersion energy in Lifshitz-like form:
\begin{align}
U^{\vdW}(D)=&\Ec(D)-\Ec(\infty)
\nonumber\\
=&\int_0^{\infty}\frac{\d\sigma}{\pi} T(\sigma)
\label{eqn:UvdW}
\\
T(\sigma)=&
\Tr\lbrs \log(1 - \chih_{1A}\hat{V}_{AB}\chih_{1B}\hat{V}_{BA}) \rbrs
\label{eqn:TvdW}
\end{align}
wbere $\chih_{1A/B}$ is the interacting response of system A/B in isolation
defined in the dRPA as
\begin{align}
\chih_{1A/B}=&\chih_{0A/B} + \chi_{0A/B}\hat{v}\chi_{1A/B}.
\end{align}
$\hat{V}_{AB}=\hat{V}_{BA}$ is the Coulomb
potential between electrons in different systems \emph{only}
and can be considered a function of $\vr_A,\vr_B$ and $D$
where $\vr_{A/B}$ is a position in system A/B.
Equations \eqref{eqn:UvdW}-\eqref{eqn:TvdW} follow
from analysis of Feynmann ring diagrams
(see \cite{Despoja2006} for a similar analysis)
or from the ACFD where the intra-system Coulomb interaction
$\hat{v}$ and $\hat{V}_{AB}$ are switched
on separately.

Let us define a system composed of two periodic slabs
localised in $z$ (ie. where we can define a length $s$ such that
$\chi_{1A/B\vq}(z,z')$ is negligible for $|z|\geq s/2$ or $|z'|\geq s/2$)
and centered such that $\int z \d z \n_{A/B}(z)=0$. In such a system
it can be shown\footnote{A full proof is beyond the scope of this
supplement but can be provided on request. It arises from a cancellation
of most terms, in the limit $D\to\infty$,
due to the dominance of $|\vq_{\pa}|=O(1/D)$. This means that
$e^{-|\vq_{\pa}+\vG_{\pa}|D}\ll e^{-q_{\pa}D}$ for $\vG_{\pa}\neq 0$
and $\int\d z X(z)e^{|q_{\pa}|z}\aeq\int\d z X(z)$} that,
in the limit $D\gg s$ and $D\to\infty$, \eqref{eqn:TvdW} is equivalent to
\begin{align}
T(\sigma)\aeq&
\int_{\BZ}\d\vq_{\pa} \lbrs \log(1 - \bar{X}_{1A\vq_{\pa}}\bar{X}_{1B\vq_{\pa}}
q_{\pa}^2e^{-2q_{\pa}D}) \rbrs.
\label{eqn:TvdWAs}
\end{align}
Here $\bar{X}_{1A/B\vq_{\pa}}(i\sigma)=\int\d z\int\d z'
(\vech{q}_{\pa}-i\hat{z})\cdot\ten{X}_{1A/B\vq_{\pa}}(\vec{0},z,\vec{0},z';i\sigma)
\cdot(\vech{q}_{\pa}+i\hat{z})$
where we project the tensor response
$\ten{X}_{1A/B\vq}(\vG_{\pa},z,\vG_{\pa}',z';i\sigma)$
in reciprocal space in the $xy$-plane and real space in $z$.

In an insulating system we have shown that
$|\tenh{X}_{1\vq}(i\sigma)|\leq \frac{Y}{W^2+\sigma^2}$ and thus
$|\bar{X}_{1\vq}(i\sigma)|\leq \frac{Y'}{W^2+\sigma^2}$
where $Y$ and thus $Y'$ is finite. Since $W$ is finite there
is a well-defined upper bound for insulating systems and
we need not worry about singularities in
$\bar{X}_{1A/B\vq}$ we can set $x=\qpa D$ and
make a series expansion in $D^{-1}$ such that
\begin{align}
T(\sigma)\aeq & 
\frac{K}{D^2}\int_0^{k_{F\pa}D} 2\pi x\d x
\log[1-\bar{X}_{1A\vec{0}}\bar{X}_{1B\vec{0}}\frac{x^2}{D^2}e^{-x}]
\nonumber\\
\aeq& -\frac{K'}{D^4}\bar{X}_{1A\vec{0}}\bar{X}_{1B\vec{0}} + O(D^{-6}).
\end{align}
Inserting this into \eqref{eqn:TvdWAs} we find
\begin{align}
U^{\vdW}(D)=& -\frac{K'}{D^4}\int_0^{\infty}\frac{\d\sigma}{\pi}
\bar{X}_{1A\vec{0}}(i\sigma)\bar{X}_{1B\vec{0}}(i\sigma)
\\
=&-\frac{C_4}{D^4}.
\end{align}
Ergo two insulating slabs have a $D^{-4}$ power law in CM theory
as in full dRPA. This conclusion applies only for insulators:
for thin metals the $\sigma$ integration diverges.

\section{Two-slab geometry}

Let us define our two-slab metal problem to have a background charge
$n^+(z)=\rho[H(\frac{s}{2}-|z-L|) + H(\frac{s}{2}-|z+L|)]$
where $L=\frac{(D+s)}{2}$ and $H(x)=1\forall x\geq 0,0$ otherwise.
This defines two jellium slabs of width $s$,
surface-to-surface distance $D$ and backround charge per unit area
$\rho=3/(4\pi r_s^3)$. The total number of electrons per unit area
is set to $N_s=2s\rho=\int_{-\infty}^{\infty}n^+(z)$.

The partial isotropy means $V^{\KS}(\vr)\equiv V^{\KS}(z)$
and the KS wavefunctions take the form
\begin{align}
\psi_{n\vkpa}(\vr)=&p_n(z)e^{-i\vkpa\cdot\vrpa}
\end{align}
where $\int\d z p^*_n(z)p_m(z)=(2\pi)^{-2}\delta_{nm}$.
The KS energies are $\epsilon_{n\vkpa}=\epsilon_{n\vec{0}}+\half |\vkpa|^2$
with occupation $f_{n}=2\max(\epsilon_F-\epsilon_{n\vec{0}},0)$.
The density and kinetic pressure tensor are thus
\begin{align}
\n(z)=&\sum_n f_{n}|p_n(z)|^2
\\
\tT(z)=&\tpa(z)[\xh\otimes\xh+\yh\otimes\yh] + \tz(z)\zh\otimes\zh
\end{align}
where $\tpa(z)=\sum_n f_{n}
\frac{\epsilon_F-\epsilon_{n\vec{0}}}{2}|p_n(z)|^2$ and
$\tz(z)=\sum_n f_{n} |\Dp{z}p_n(z)|^2-\frac{1}{4}\Dp{zz}\n(z)$.

For the present slab problem we choose auxiliary basis functions
of the form $\phi_{k\vqpa}(\vr)=b_k(z)e^{-i\vqpa\cdot\vrpa}$
where $b_k(z)$ is either $e^{in\frac{\pi z}{s}}$ or
$\tanh(k_tz)e^{in\frac{\pi z}{s}}$ where $k_t$ is a parameter chosen
to optimise convergence and $n$ is an integer.
The restiction to integer $n$ makes this basis set
incomplete but inclusion of non-integer $n$ does not alter results.

We then set 
\begin{align}
\vu_{N\vqpa}=&\sum_k \phi_{k\vqpa}(\vr)
[a^k_{Nz}(\qpa)\zh + a^k_{N\pa}(\qpa)\vhqpa]
\end{align}
(the $\hat{\vq}_{\perp}=\vqpa\times\zh$ term does not
contribute to the correlation energy). Thus the eigen-equations are
\begin{align}
\Omega_N^2(\qpa) \N_{jk} a^k_{N\pa}(\qpa)
=&K_{jk\pa\pa}(\qpa) a^k_{N\pa}(\qpa) 
\nonumber\\&
+ K_{jk\pa z}(\qpa) a^k_{Nz}(\qpa)
\allowdisplaybreaks\\
\Omega_N^2(\qpa) \N_{jk} a^k_{Nz}(\qpa)
=&[\Phi^0_{jkzz} + K_{jk zz}(\qpa)] a^k_{Nz}(\qpa)
\nonumber\\&
+ K_{jk z\pa}(\qpa) a^k_{N\pa}(\qpa)
\end{align}
which must be solved for each $\qpa$. Normalisation
gives $\sum_{jk} \N_{jk} [a^{j*}_{N\pa}a^k_{M\pa} + a^{j*}_{Nz}a^k_{Mz}]=
(2\pi)^{-2}\delta_{NM}$.

Here $N_{jk}=\int\d z \n(z)b_j^*(z)b_k(z)$ and
$\Phi^0_{jkzz}=\int\d z \n(z)[\Dp{zz} V^{\KS}(z)]b_j^*(z)b_k(z)$
are independent of $\qpa$.
The components of $\mat{K}(\qpa)$ take the form
\begin{align}
-K_{jk zz}
=&3\FF_{jk}[\tz,1,1]
+\frac{1}{4} \FF_{jk}[\n,2,2]
\nonumber\\&
 + \qpa^2\lbr \FF_{jk}[\tz,0,0]
+ \frac{1}{4}\FF_{jk}[\n,1,1]
\rbr
\label{eqn:KForm1}
\\\allowbreak
-K_{jk\pa\pa}
=&\FF_{jk}[\tpa,1,1]
+ \frac{\qpa^4}{4}\FF_{jk}[\n,0,0]
\nonumber\\&
+ \qpa^2\lbr 3\FF_{jk}[\tpa,0,0] 
+ \frac{1}{4}\FF_{jk}[\n,1,1] \rbr
\label{eqn:KForm2}
\\\allowbreak
-K_{jk\pa z}
=&(-i\qpa)\lbr \FF_{jk}[\tz+\tpa,0,1]
+ \frac{1}{4} \FF_{jk}[\n,1,2] \rbr
\nonumber\\&
+ \frac{-i\qpa^3}{4} \FF_{jk}[\n,0,1].
\label{eqn:KForm3}
\end{align}
where we use the shorthand
$\FF_{jk}[f,a,b]= \int f(z) [\Dp{z^a}b_j^*(z)][\Dp{z^b}b_k(z)] \d z$.

Finally in this basis
\begin{align}
w_{N\qpa}(q_z)=&iv^{1/2}(\sqrt{\qpa^2+q_z^2})
\nonumber\\&\times
\int\d z e^{iq_zz}[\qpa u_{N\vqpa\pa} + q_z u_{N\vqpa z}],
\\
W_{NM}(\qpa)=&-\int\frac{\d q_z}{2\pi}w_{N\qpa}^*(q_z)w_{M\qpa}(q_z).
\end{align}
and
\begin{align}
\Ec^{\lr\TVF{\mu}}=&-\int\frac{\d\sigma}{2\pi}
\int\frac{2\pi\qpa\d\qpa}{(2\pi)^2}
\Tr[L(\mat{B}(\qpa,i\sigma))]
\\
B_{NM}=&\sqrt{f_N(\sigma)f_M(\sigma)}W_{NM}(\qpa)
\end{align}
where $L(x)=\log(1+x)-x$.

In our calculations we use approximately 500-1000 regularly
distributed $z$ points for quadrature (with the number
depending on system size). We also use approximately 500 $q_z$
points on a Gauss-Hermite grid (due to the Range-Separation
term $e^{-q^2/(2q_{\rm{RS}})^2}$
to calculate $W_{NM}(\qpa)$. This is more than sufficient
to represent the chosen basis functions in either space.

To correctly integrate over frequency we require a grid that
accurately deals with functions of form $a/(b^2+\sigma^2)$ where
$b$ ranges from very small to large. Choosing a regular
grid for $\sigma\ll 1$ and using a
Clenshaw-Curtis grid for larger $\sigma$ seems to work well for
these problems.

\bibliography{vanDerWaals,ACFDT,DFT,Wannier,Misc,Experiment,Hybrid,CM-RPA}